\documentclass[%
 reprint,
 amsmath,amssymb,
 aps,
]{revtex4-1}

\usepackage{graphicx}
\usepackage{dcolumn}
\usepackage{bm}

\begin{document}


\title{Symmetry consideration in the problem of wave modes\\ of thin viscous liquid layer flow}
\thanks{This work was financially supported by the Russian Scientific Foundation (No.14-22-00174)}

\author{Dmitry Arkhipov}
\author{Ivan Vozhakov}
 \email{vozhakov@gmail.com}
\author{Dmitry Markovich}
\author{Oleg Tsvelodub}
\affiliation{%
 Institute of Thermophysics, Novosibirsk, Russia\\
 Novosibirsk State University, Novosibirsk Russia
}%

\date{\today}

\begin{abstract}
The equations in conservative form for nonlinear waves modeling on a liquid film flowing down a vertical plane have been investigated. It has been found that in the computational domain extended along the transverse axis the equations with boundary conditions are invariant under parity transformation. It is numerically shown that for moderate Reynolds numbers the steady-state travelling solutions of the equations have the detected symmetry. It is demonstrated that using this symmetry for the numerical solution of the problem by Galerkin methods significantly increases the efficiency of calculations. 
\begin{description}
\item[PACS numbers]
47.15.gm
\end{description}
\end{abstract}

\maketitle


\section{Introduction}

Film flows are widely used in heat and mass transfer technologies. Theoretical study of the problem started in 1916 with Nusselt’s work that has become fundamental for the theory of film condensation. The research provided the exact solution for waveless flowing of a thin layer of viscous liquid over a solid inclined wall. The natural development of waves on the free surface of a film additionally intensifies the process of heat transfer from the heated substrate. The study of the basic characteristics and the form of these waves in the pioneer works of Kapitza \cite{kapitza1949experimental} laid the foundation for a new scientific direction.

Many of the researchers (for example, \cite{ruyer1998modeling}, \cite{kalliadasis2011falling}) use low-dimensional Galerkin methods to derive reduced systems of equations for coefficients at respective basis functions, depending on the transverse coordinate. The main advantage of such approaches results from rather small number of equations (two or three) at a lucky choice of the basis. In our work we point to a specific symmetry of governing equations and a wide class of their solutions, including those of interest from experimental point of view. It allows two-fold reduction of the number of equations and gives a deep understanding of  successfulness of the known models, derived earlier by extrapolation of analytical results, obtained only for small Reynolds numbers.

The full problem formulation for waves on isothermal flowing film is clear and includes a system of Navier-Stokes and continuity equations with appropriate boundary conditions at the wall and at the free surface. This formulation implies the solution of the problem in a changing flow area unknown in advance, which greatly complicates the mathematical and numerical simulation.

\begin{figure}[h]
\includegraphics[width=\linewidth]{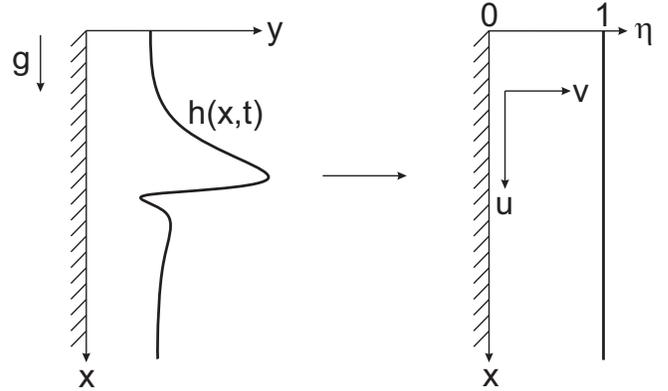}
\caption{The flow area in $x$-$y$ and $x$-$\eta$ coordinates}
\label{fig:transform}
\end{figure}

One of the approaches to solving the moving boundary problem appeared in the mid 80-ies of the last century. It comes to rewriting of hydrodynamic equations in new variables, transforming the flow area into the strip of constant thickness (Fig. \ref{fig:transform}):
\begin{equation} {\label{eq:transform}}
x = x,\quad \eta  = \frac{y}{{h(x,t)}},\quad t = t
\end{equation}
Here $h$ is an instant local film thickness. The coordinate system (\ref{eq:transform}) is non-orthogonal, so the normal vector formulation of equations is inapplicable. For this reason, many authors (see for example \cite{geshev1985hydrodynamic, trifonov2012stability}) reduce the method to a simple change of variables without transformation of vectors and tensors, contained in the original equations.

For the case of a free falling film on a vertical plane, the transformation (\ref{eq:transform}) was performed in \citep{alekseenko2011divergent} for the hydrodynamic equations written in tensor form invariant under arbitrary coordinate transformation. As a result, in the long wavelength approximation the following system was obtained:
\begin{subequations} {\label{eq:equations}}
\begin{equation}
\frac{{\partial (hu)}}{{\partial t}} + \frac{{\partial (h{u^2})}}{{\partial x}} + \frac{{\partial (huv)}}{{\partial \eta }} = \frac{\sigma }{\rho }h\frac{{{\partial ^3}h}}{{\partial {x^3}}} + \frac{\nu }{h}\frac{{{\partial ^2}u}}{{\partial {\eta ^2}}} + gh
\end{equation}
\begin{equation} {\label{eq:equations_b}}
\frac{{\partial h}}{{\partial t}} + \frac{{\partial (hu)}}{{\partial x}} + \frac{{\partial (hv)}}{{\partial \eta }} = 0
\end{equation}
\end{subequations}
Here $\sigma$ is a surface tension, $\rho$ is a density, $\nu$ is a kinematic viscosity, $g$ is a free fall acceleration, $u$ and $v$ are contravariant components of velocity corresponding to the coordinates $x$ and $\eta$, respectively. It is clear that since $x$ has remained an original Cartesian coordinate, the component $u(x,\eta,t)$ in contrast to $v(x,\eta,t)$ remains the longitudinal component of physical velocity.

Boundary conditions on the solid wall are: 
\begin{equation} {\label{eq:bound1}}
u(x,0,t) = 0, \quad v(x,0,t) = 0
\end{equation}
No-stress (\ref{eq:bound2a}) and kinematic conditions (\ref{eq:bound2b}) on the free surface are:
\begin{subequations} {\label{eq:bound2}}
\begin{equation} {\label{eq:bound2a}}
\frac{{\partial u}}{{\partial \eta }}(x,1,t) = 0
\end{equation}
\begin{equation} {\label{eq:bound2b}}
v(x,1,t) = 0
\end{equation}
\end{subequations}

In this problem there are three unknown functions, one of which $h$ is the function of only two variables $(x,t)$. In the calculated domain $(x,\eta)$ there are 2 unknown functions -- $u(x,\eta,t)$ and $v(x,\eta,t)$, therefore, 2 equations are necessary to find them. In the long-wave model these two equations are: continuity equation and momentum balance in the longitudinal direction. A kinematic condition on the free surface $(\eta=1)$, allowing finding the function $h(x,t)$. In contrast to usual presentation of the kinematic condition on a free boundary in Cartesian coordinates, where it is represented as a differential equation:

$$v(x,y=h,t)=\frac{\partial h}{\partial t} + u(x,y=h,t)\frac{\partial h}{\partial x}$$ 

in curvilinear coordinates (1) it takes the form (\ref{eq:bound2b}):

\section{Symmetry}

Let new transverse coordinate be
$$
\eta' = \eta  - 1
$$
so that the flow area lies in the interval $\eta' \in [-1,0]$. Notice that $\eta'$ is expressed in the original Cartesian coordinates as follows:
$$
\eta' = \frac{y}{h(x,t)} - 1
$$
It is easy to see that the equations (\ref{eq:equations}) are invariant under the transformation:

\begin{subequations} {\label{eq:symmetry}}
\begin{equation}
\eta' \to  -\eta'
\end{equation}
\begin{equation}
u(x,\eta ',t) \to u(x, - \eta ',t)
\end{equation}
\begin{equation}
v(x,\eta ',t) \to  - v(x, - \eta ',t)
\end{equation}
\end{subequations}

It means in particular that there are two types of solutions of these equations in the extended strip $\eta' \in [-1,1]$. Solutions of the first type are characterized by symmetry:
\begin{subequations} {\label{eq:sym_bound1}}
\begin{equation}
u(x,\eta ',t) = u(x,-\eta',t) 
\end{equation}
\begin{equation}
v(x,\eta',t) =  - v(x,-\eta ',t)
\end{equation}
\end{subequations}

It is clear that the solutions of the first type satisfying the no-slip and no-penetration boundary conditions on both boundaries:
\begin{subequations} {\label{eq:sym_bound2}}
\begin{equation}
u(x, - 1,t) = v(x, - 1,t) = 0
\end{equation}
\begin{equation} {\label{eq:sym_bound2_b}}
u(x,1,t) = v(x,1,t) = 0
\end{equation}
\end{subequations}
are the solutions to original problem (\ref{eq:equations})-(\ref{eq:bound2}) in a half-strip $[-1,0]$. Indeed on the boundary $\eta'=-1$ the conditions (\ref{eq:bound1}) are performed, and the boundary condition of no-penetration at $\eta'=0$ (kinematic condition – $v(x,0,t) = 0$) is satisfied automatically, due to fact that the contravariant transverse component of the velocity $v$ for the first type solutions is an odd function with respect to $\eta'$. And since the function $u$ is even, the dynamic condition on this boundary is automatically performed as well:
$$
\frac{{\partial u}}{{\partial \eta '}}(x,0,t) = 0
$$

Solutions of the second type do not have such symmetry, but in virtue of (\ref{eq:symmetry}), if there is a solution $u_1(x,\eta ',t)$ and $v_1(x,\eta ',t)$, then there is a solution:
$$
u_2(x,\eta',t) = u_1(x,-\eta',t)
$$
$$
v_2(x,\eta',t) =-v_1(x,-\eta ',t) 
$$
At that
$$
u_1(x,\eta',t) \ne u_1(x,-\eta',t)
$$
$$
v_1(x,\eta',t) \ne  - v_1(x,-\eta',t)
$$

However, if the second type solutions for the problem (\ref{eq:equations})-(\ref{eq:bound2}) are extended into interval $\eta' \in [0,1]$ then, boundary conditions (\ref{eq:sym_bound2_b}) at $\eta'=1$ are not to be performed.

Considering higher orders of smallness makes the problem as a whole and boundary conditions on free surface (\ref{eq:bound2}) in particular much more complicated. All this will most probably lead to symmetry (\ref{eq:symmetry}) breaking. Anyway, the pointed-out symmetry will be destroyed even at the used approximation if other conditions at free boundary, for example, the ones taking into account gas influence on film flowing, are considered. Nevertheless, the found symmetries for the model of a boundary layer, are quite interesting themselves.

The above symmetry properties of equations may be useful for finding solutions to the problem (\ref{eq:equations})-(\ref{eq:bound2}).

\section{Calculations}

For numerical solution the problem (\ref{eq:equations})-(\ref{eq:bound2}) was written in a dimensionless form:
\begin{subequations} {\label{eq:equat_QV}}
\begin{eqnarray}
\varepsilon Re\left( \frac{\partial Q}{\partial t} + \frac{\partial }{\partial x}\left( \frac{Q^2}{h} \right) + \frac{\partial }{\partial \eta '}\left( \frac{QV}{h} \right) \right) = \nonumber \\
 = \frac{1}{h^2}\frac{\partial ^2Q}{\partial \eta'^2} + 3h + \frac{3^{1/3}Fi^{1/3}}{Re^{5/3}} \varepsilon ^3 Re h\frac{\partial ^3 h}{\partial x^3} {\label{eq:equat_QV_a}}
\end{eqnarray}
\begin{equation}
\frac{{\partial h}}{{\partial t}} + \frac{{\partial Q}}{{\partial x}} + \frac{{\partial V}}{{\partial \eta '}} = 0
\end{equation} 
\begin{equation}
Q(x,-1,t) = V(x,-1,t) = 0
\end{equation}
\begin{equation}
\frac{{\partial Q}}{{\partial \eta '}} (x,0,t) = 0, \quad V(x,0,t) = 0
\end{equation}
\end{subequations}
We introduce new functions $Q=uh$, $V=vh$ relative to which the equation (\ref{eq:equations_b}) is linear.  Here $Re$ is Reynolds number, $Fi=\sigma^3 / \rho^3 g \nu^{4/3}$ is a film number characterizing physical properties of liquid, $\varepsilon$ is a long-wave parameter, $u_0$, $l_0$, $h_0$ and $l_0/u_0$ are characteristic scales of velocity, length, thickness and time, respectively.
The solution of the linear problem (\ref{eq:equations})-(\ref{eq:bound2}) shows that the unperturbed flow
$$
h_0 = 1, \quad Q_0(\eta') = \frac{3}{2}(1 - \eta'^2), \quad V_0 = 0
$$
is unstable with respect to linear perturbations of the type 
$$
\left[Q',h',V' \right] = \left[Q_a',1,V_a' \right]\,\left(\eta'\right)h_a'\exp(i\alpha(x - ct)) + c.c.
$$
for an interval of wave numbers $\alpha$ less than $\alpha_n$. Here $c=c_r + i c_i$ is complex phase velocity.
The scales of length and thickness (and hence the parameter $\varepsilon$) may be selected so that the wave number of neutral perturbations ($c_i=0$) for any values of Reynolds numbers was equal to $\alpha_n=1$. With this normalization for the undisturbed flow the area of instability ($c_i>0$) with respect to linear perturbations would take an interval $0 < \alpha  < \alpha_n = 1$. Accordingly, the wave numbers of stable perturbations would be in the area $\alpha  > \alpha_n = 1$. For such normalization it is necessary to determine the relationship between parameters $\varepsilon, Re, Fi$ from the solution to a linear problem. For the problem (\ref{eq:equations})-(\ref{eq:bound2}) such relationship in the general case cannot be determined analytically.

\begin{figure}[h!]
\begin{minipage}[h!]{0.47\linewidth}
\center{\includegraphics[width=1\linewidth]{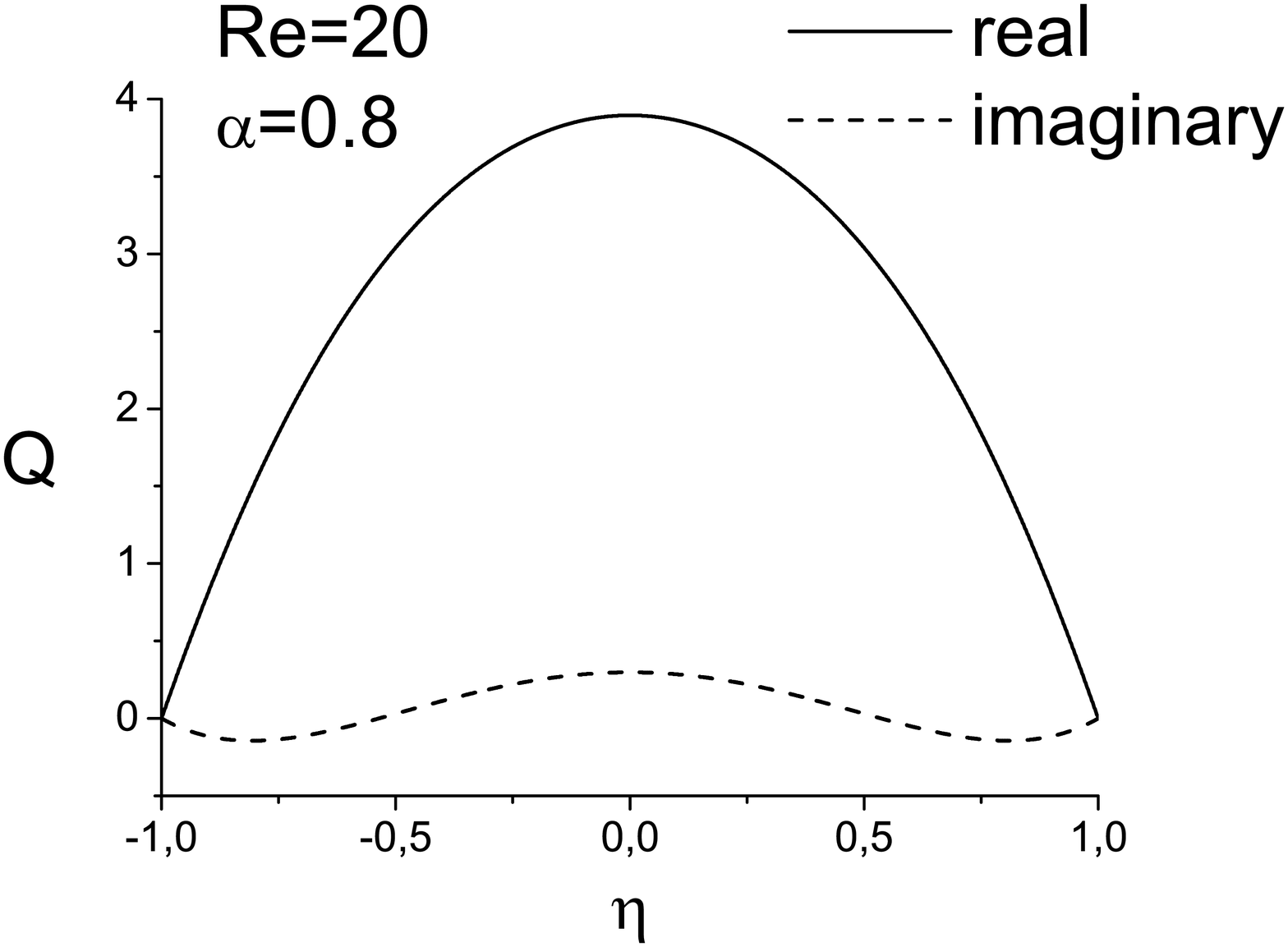}} \\
\end{minipage}
\hfill
\begin{minipage}[h!]{0.47\linewidth}
\center{\includegraphics[width=1\linewidth]{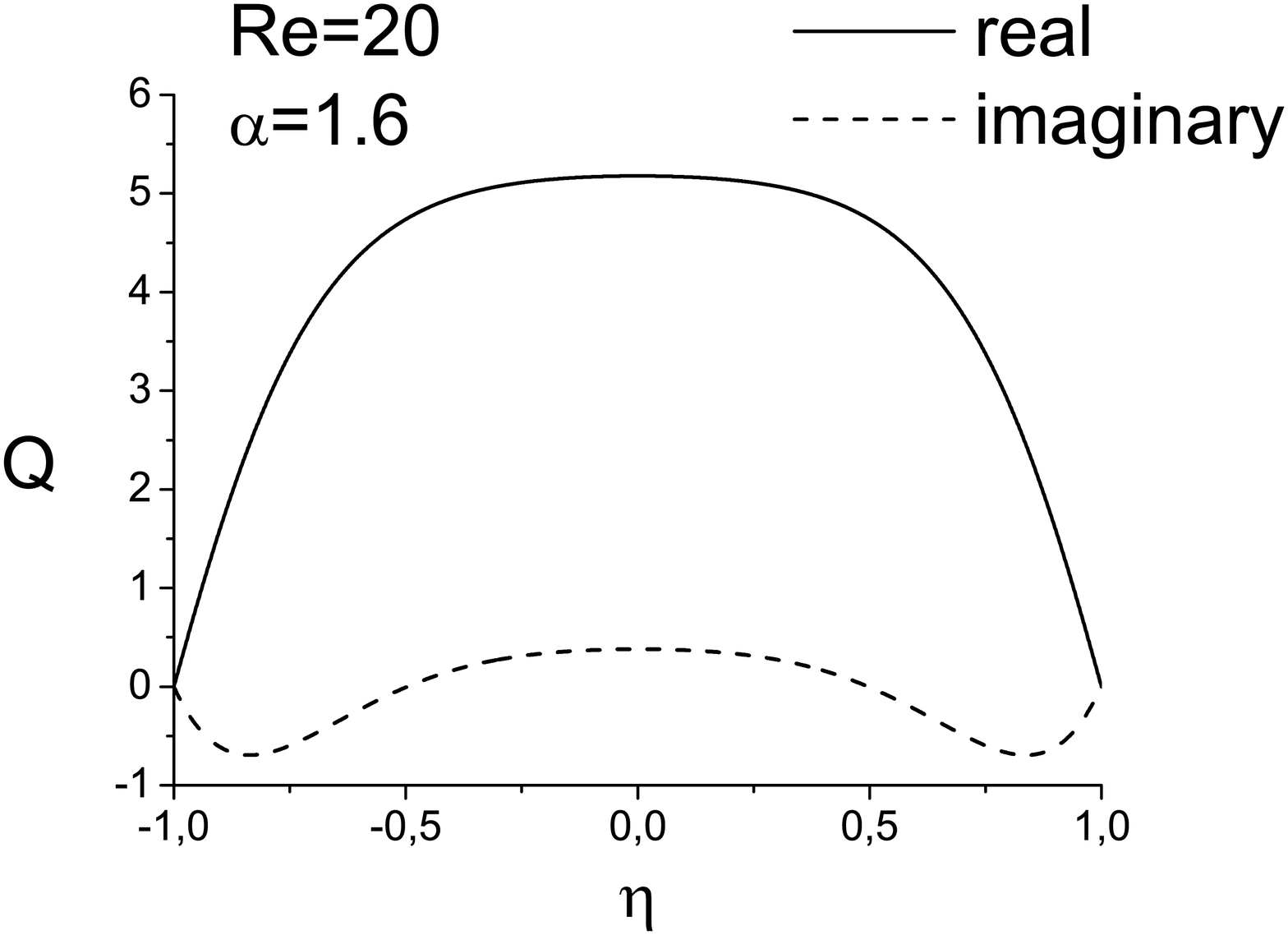}} \\
\end{minipage}
\vfill
\begin{minipage}[h!]{0.47\linewidth}
\center{\includegraphics[width=1\linewidth]{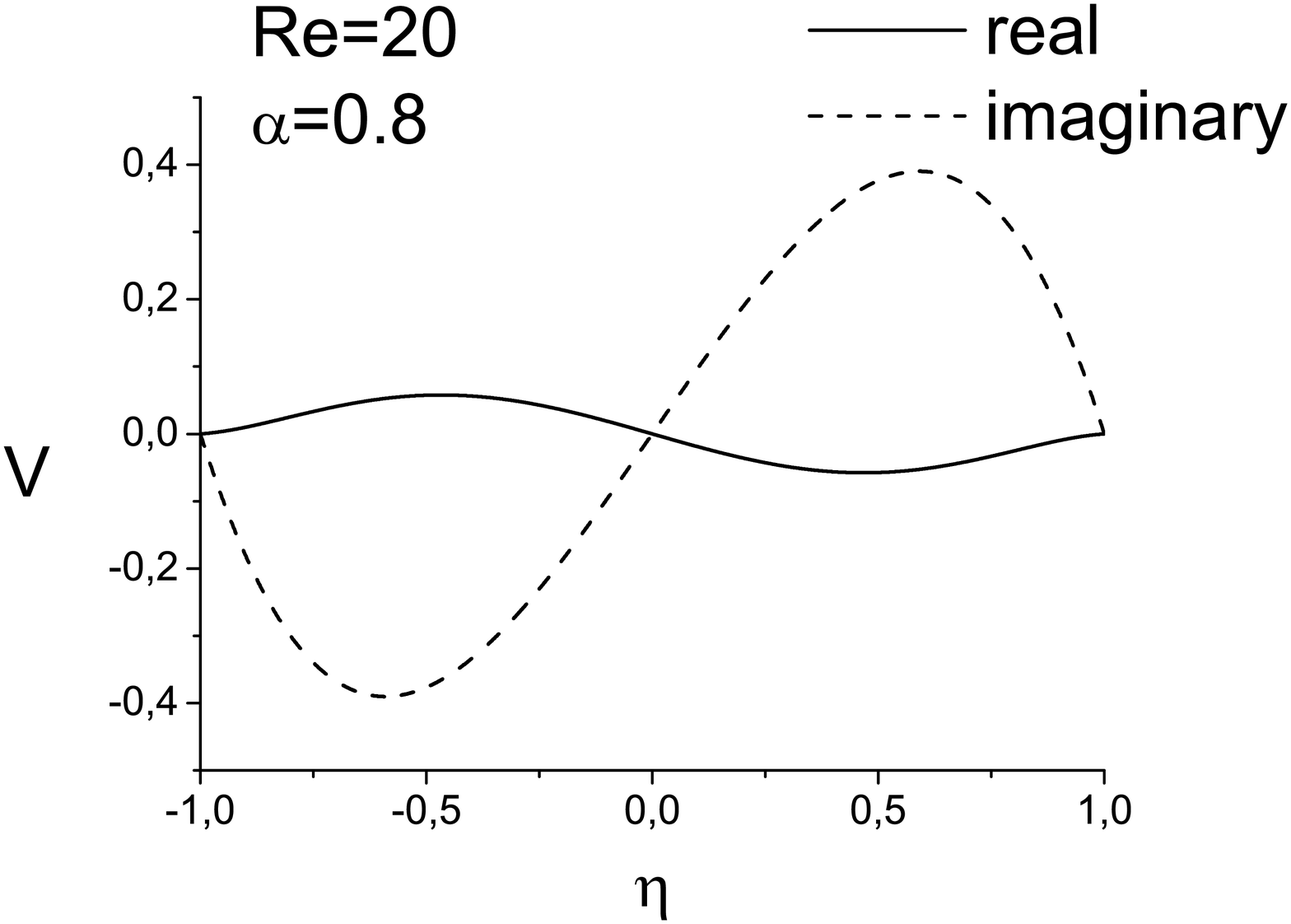}} \\
\end{minipage}
\hfill
\begin{minipage}[h!]{0.47\linewidth}
\center{\includegraphics[width=1\linewidth]{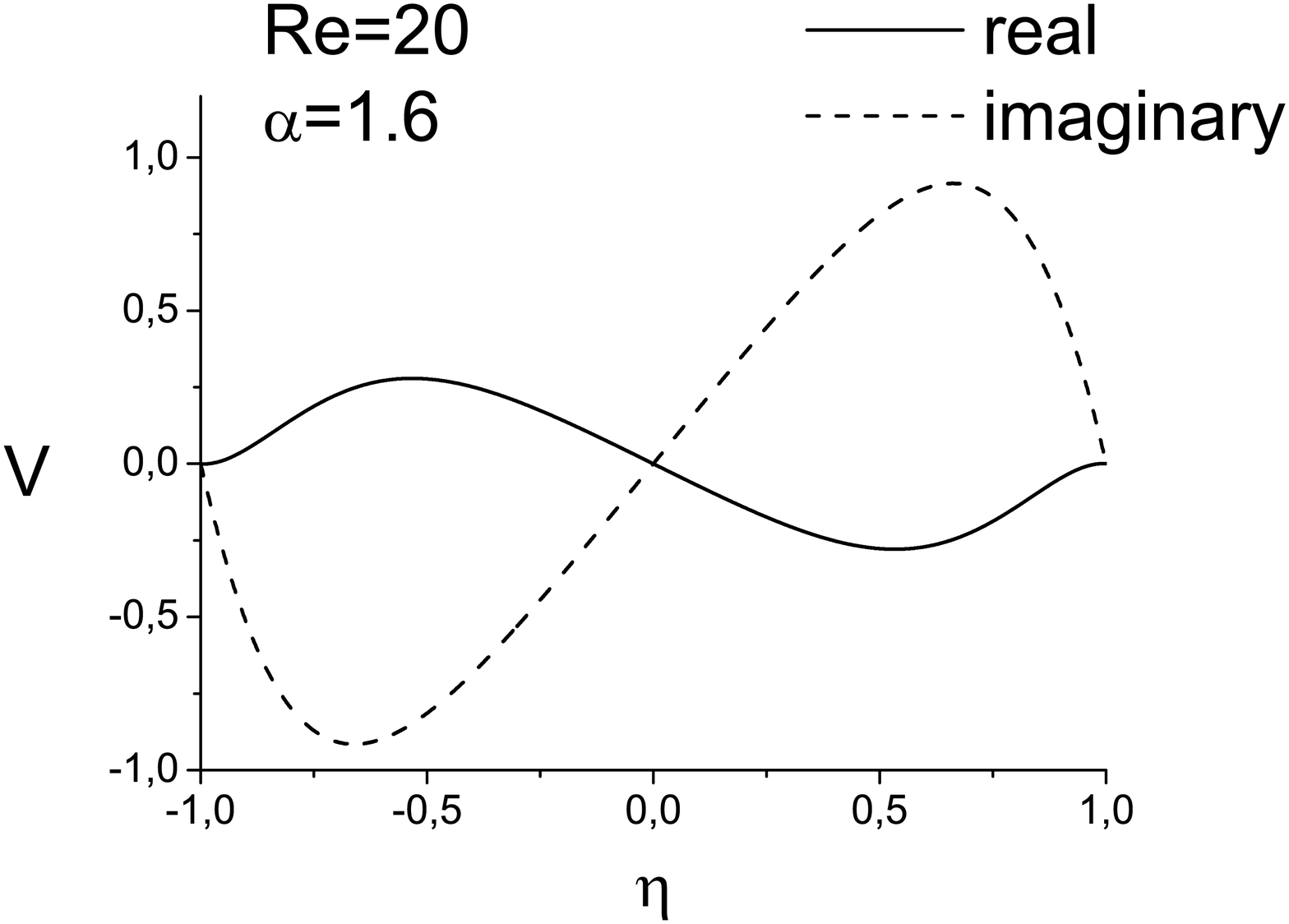}} \\
\end{minipage}
\caption{Real ($Q_r$,$V_r$) and imaginary ($Q_{im}$,$V_{im}$) parts of $Q$ and $V$ depending on dimensionless transverse coordinate in extended flow area.}
\label{fig:linear}
\end{figure}

In \cite{arkhipov2011investigation} it is shown that at low Reynolds numbers, the problem (\ref{eq:equations})-(\ref{eq:bound2}) are reduced to the equation of Kuramoto-Sivashinsky (Nepomniashchi), where the wave number of neutral perturbations is known (see for example \cite{arkhipov2012comparison}:
$$
{\varepsilon ^2}\alpha_n^2 = \frac{6}{5}\frac{{3{{{\mathop{\rm Re}\nolimits} }^{5/3}}}}{{{3^{1/3}}F{i^{1/3}}}}
$$
Hence, assuming $\alpha_n=1$, for $\varepsilon$ obtain:
\begin{equation} {\label{eq:epsilon}}
\varepsilon  = \sqrt {\frac{6}{5}\frac{{3{{{\mathop{\rm Re}\nolimits} }^{5/3}}}}{{{3^{1/3}}F{i^{1/3}}}}}
\end{equation} 
This is the ratio that was used to consider the problem (\ref{eq:equat_QV}) in the area of moderate Reynolds numbers. In accordance with calculations, up to values $\varepsilon Re = 6$ (for water films is $Re \approx 20$) $\alpha_n$ differs from unity by less than 0.002. Taking into account expression (\ref{eq:epsilon}), the coefficient of the capillary term in the equation (\ref{eq:equat_QV_a}) takes the form: 
$$
\frac{3^{1/3}F^{1/3}}{Re^{5/3}} \varepsilon^3 Re=\frac{18}{5} \varepsilon Re
$$
so only parameter $\varepsilon Re$ remains in the problem. 
To solve a linear problem, the algorithm described in \cite{arkhipov2012comparison} was used: the equations (\ref{eq:equat_QV}) was reduced to one equation of the modified stream function, then the equation was linearized and solved by the shooting method from the substrate $\eta'=-1$ (to solve the Cauchy problem we used the 4th-order Runge-Kutta method). The dynamic condition on the free boundary determined a free coefficient in a homogeneous solution to the equation. The phase velocity was found by Newton's method from the kinematic condition on the surface.
According to direct numerical calculations, extending the linear solution to the problem (\ref{eq:equat_QV}) into the interval $\eta' \in [0,1]$ we obtain solutions with symmetry both for stable and unstable perturbations. Two examples of such calculations are shown in Fig. \ref{fig:linear}.

The growth of unstable linear perturbations may be eventually stopped by the action of nonlinear effects, resulting in the formation of steady-state non-linear regimes. Therefore, in the studies of wave flows of liquid films much attention is paid to the steady-state travelling waves. In this work we will limit consideration to periodic steady-state travelling solutions of system (\ref{eq:equat_QV}): 
$$
\left[ {Q,h,V} \right] = \left[ {Q,h,V} \right]\,\left( {\tilde x,\eta '} \right),\quad \tilde x = \alpha (x - ct)
$$

In connection with the results obtained for linear solutions, the following question arises, how wide is the class of steady-state travelling solutions of the problem (\ref{eq:equations})-(\ref{eq:bound2}), for which the symmetry (\ref{eq:sym_bound1}) is performed as well?

Currently the interest in Galerkin spectral methods as applied to the problems of fluid motion has markedly increased. To describe film flows a number of new model systems obtained by this method were introduced. The minimum number of members in Galerkin’s approaches, required for a good approximation of solutions, greatly depends on how well the basis functions were chosen. 
For the numerical study we used the pseudospectral method, in which functions depending on the transverse coordinate $\eta'$ were expanded in series on Chebyshev polynomials $T_i$:
\begin{subequations} {\label{eq:polinoms}}
\begin{equation}
Q = \sum {Q_i(x) T_i(\eta')}
\end{equation}
\begin{equation}
V = \sum {V_i(x) T_i(\eta')}
\end{equation}
\end{subequations}
and functions depending on the longitudinal coordinate were expanded in Fourier series:
\begin{subequations} {\label{eq:harmonics}}
\begin{equation}
Q_i(\tilde x) = \sum {Q^k_i e^{ik\tilde x}}
\end{equation}
\begin{equation}
V_i(\tilde x) = \sum {V^k_i e^{ik\tilde x}}
\end{equation}
\begin{equation}
h(\tilde x) = \sum {H^k e^{ik\tilde x}}
\end{equation}
\end{subequations}

Limiting to the first N harmonics in (\ref{eq:harmonics}) and substituting expressions (\ref{eq:polinoms})-(\ref{eq:harmonics}) in system (\ref{eq:equat_QV}) written for a variety of Chebyshev nodes:
$$
{\eta'_j} = \cos \left( {\frac{{2j - 1}}{{2M}}\pi } \right),\,j = 1..M
$$
and satisfying the boundary conditions at the points $\eta'=-1$ and $\eta'=0$, we come to a finite-dimensional system of nonlinear algebraic equations for coefficients $Q_i^k, V_i^k, H^k, c$.
The resulting problem was solved by the iterative method of Newton-Kantorovich: 
$$
\left[ {Q,h,V,c,\alpha } \right]_{n + 1} = \left[ {Q,h,V,c,\alpha } \right]_n + \left[ {\delta Q,\delta h,\delta V,\delta c,\delta \alpha } \right]
$$
Here the symbol $\delta$ indicates the variation of the corresponding value. After linearization relative to variations the problem is reduced to a system of linear algebraic equations.

Using algorithm described above a series of calculations of steady-state travelling solutions of the system (\ref{eq:equat_QV}) in the strip $\eta' \in [-1,1]$ was performed. The parameter $\varepsilon Re$ varied in the range $[0.03 \div 8]$. For the case of water this means a change of Reynolds number in the range $[1 \div 30]$. A substantial part of the calculations was performed on the bases (\ref{eq:polinoms}) with the number of polynomials $T_i$ -- 9, 11, 13. The calculations have shown that the difference in the waveform obtained on the bases (\ref{eq:polinoms}), containing 9 and 11 polynomials, was discernible in areas with large gradients of thickness (on the wave front and in the field of precursor). But even in these areas the results, built on the bases of 11 and 13 polynomials, coincided with graphic accuracy. The most remarkable result is that the values of coefficients obtained during calculations for all odd Chebyshev polynomials turned out to be equal to null with machine precision. Thus, it has become clear that all found steady-state travelling solutions of system (\ref{eq:equations})-(\ref{eq:bound2}) are characterized by symmetry (\ref{eq:sym_bound1}).

Considering these results, for comparison we have applied an alternative method of solving the system (\ref{eq:equat_QV}), where for the function $Q$ we used only the even Chebyshev polynomials, and for the function $V$ -- only the odd ones:
\begin{subequations} {\label{eq:even_odd_polinoms}}
\begin{equation}
Q = \sum {{Q_{2i}}\left( {\tilde x} \right)} \,\left( {{T_{2i}}\left( {\eta '} \right) - 1} \right)
\end{equation}
\begin{equation}
V = \sum {{V_{2i + 1}}\left( {\tilde x} \right)} \,\left( {{T_{2i + 1}}\left( {\eta '} \right) - \eta '} \right)
\end{equation}
\end{subequations}
Such a choice provides automatic performance of boundary conditions at the interface and on the solid wall. Of course now it is sufficient to consider points $\eta'$ only in a half-band $[-1,0]$. This representation gives the solution to our problem (\ref{eq:equations})-(\ref{eq:bound2}).
Several control calculations carried out in the basis (\ref{eq:even_odd_polinoms}), consisting of 3 even Chebyshev polynomials for the longitudinal velocity, have shown excellent agreement with the results, obtained on the full basis (\ref{eq:polinoms}) from 9 polynomials.

\begin{figure}[h!]
\begin{minipage}[h!]{\linewidth}
\center{\includegraphics[width=1\linewidth]{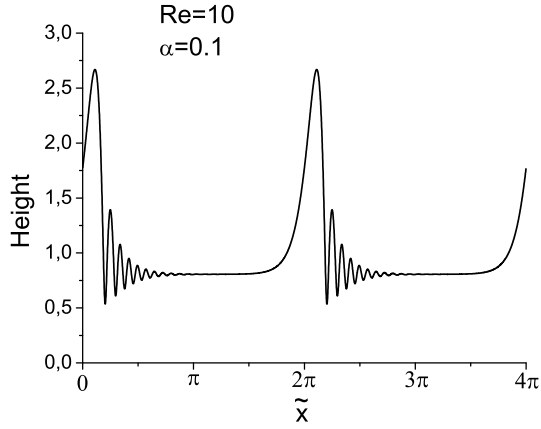}} \\
\end{minipage}
\vfill
\begin{minipage}[h!]{\linewidth}
\center{\includegraphics[width=1\linewidth]{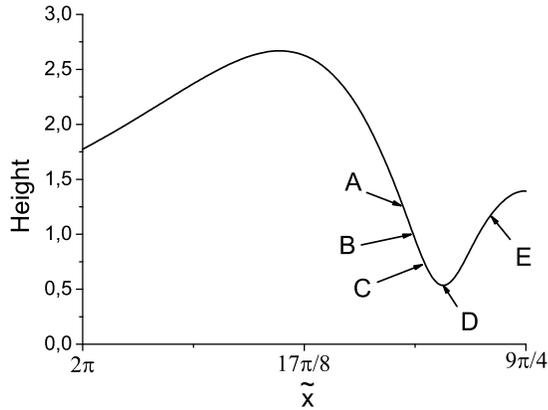}} \\
\end{minipage}
\caption{Profile of the wave}
\label{fig:wave}
\end{figure}

\begin{figure}[h!]
\includegraphics[width=\linewidth]{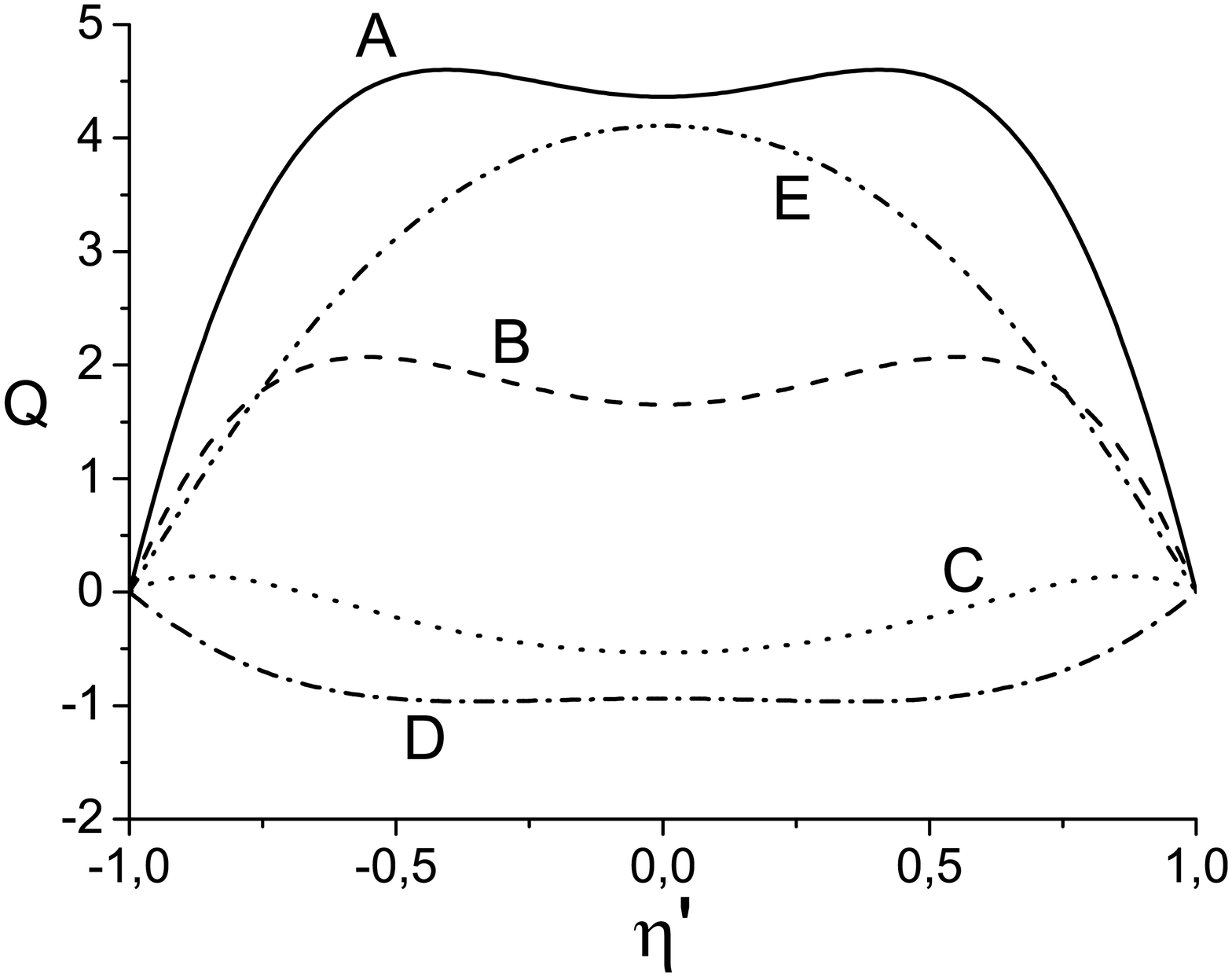}
\caption{Profiles of the function $Q$ for selected wave sections A - E}
\label{fig:Q}
\end{figure}

\begin{figure}[h!]
\includegraphics[width=\linewidth]{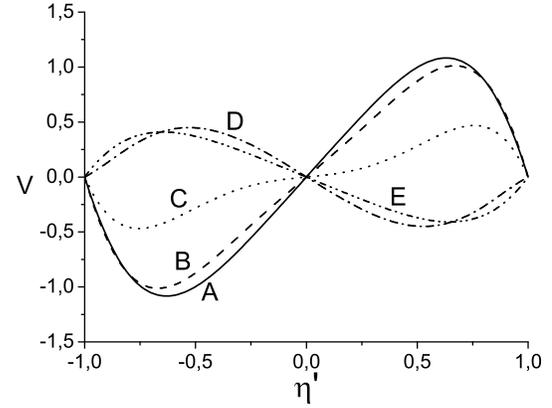}
\caption{Profiles of the function $V$ for selected wave sections A - E}
\label{fig:V}
\end{figure}

The obtained steady-state travelling solutions of the problem (\ref{eq:equations})-(\ref{eq:bound2}) are adequately presented in \cite{arkhipov2014symmetry}. Here we give one of these solutions as an example. Fig. \ref{fig:wave} presents the form of the surface for this case. For a more detailed description of the solution, it is useful to present profiles of functions $Q$ and $V$ in different wave sections. The most interesting are areas with large thickness gradients (leading wave front and precursor). Here, for example, the profile $Q$ is significantly different from the parabolic, and it may contain a point of inflection. In these areas reverse flows may occur as well. Sections in enlarged part of Fig. \ref{fig:wave} suitable from this point of view are numbered with points A-E. The corresponding profiles of functions $Q$ and $V$ are shown in Fig. \ref{fig:Q} and Fig. \ref{fig:V}. To demonstrate symmetries (\ref{eq:sym_bound1}) profiles are given in the interval $\eta' \in [-1,1]$.

Of course, on the basis of performed calculations it cannot be considered strictly proved that all possible steady-state travelling solutions of the problem (\ref{eq:equations})-(\ref{eq:bound2}) in the band $\eta' \in [-1,1]$ have symmetry (\ref{eq:sym_bound1}), but such a conclusion is very likely. At least for a special case of periodic steady-travelling modes having positive solitons as limits, which is important from the point of view of the experiment, we can assume that this is true.
Therefore, we can conclude about the usability of shorter and convenient basis functions, based on symmetry (\ref{eq:sym_bound1}) of the equations (\ref{eq:equations}), for calculating the steady-state travelling modes of wave motion of the film.

For example, the symmetry is specific for the function, used in the well-known model of Shkadov. In this context let us mention the model, used in the well known series of articles by Ruyer-Quil and Manneville \cite{ruyer1998modeling, ruyer2000improved} and show that there the basic functions are also symmetric.
To represent the longitudinal velocity $u$ the authors used a basis consisting of three polynomials:
$$g_0 = \eta  - \frac{1}{2} \eta^2$$
$$g_1 = \eta  - \frac{17}{6} \eta^2 + \frac{7}{3}\eta^3 - \frac{7}{12}\eta^4$$
$$g_2 = \eta  - \frac{13}{2} \eta^2 + \frac{57}{4}\eta^3 - \frac{111}{8}\eta^4 + \frac{99}{16}\eta^5 - \frac{33}{32} \eta^6$$

The choice of such a basis was almost pushed by the analytical results, obtained in the case of weakly-nonlinear perturbations at small flow rates, when the Reynolds numbers are of the order of unity. In this basis the velocity   is written as follows: 
$$
u(x,\eta ,t) = \sum\limits_{j = 0}^2  {b_j}(x,t){g_j}(\eta )
$$

By means of shift transforming $\eta  \to \eta'$ the basis $g_j$ of Ruyer-Quil, Manneville takes the following form:
$$ g_0 =  \frac{1}{2} - \frac{1}{2}{\eta'^2} $$
$$ g_1 = - \frac{1}{12} + \frac{2}{3}{\eta'^2} - \frac{7}{12}{\eta'^4}  $$
$$ g_2 =  \frac{1}{32} - \frac{{19}}{{32}}{\eta'^2} + \frac{{51}}{{32}}{\eta'^4} - \frac{{33}}{{32}}{\eta'^6} $$

Now it has become clear that the polynomials $g_j$ contain only even terms relative to the coordinate $\eta'$. Therefore, all possible expressions of the longitudinal velocity presented with this set will be symmetric with respect to the coordinate $\eta'$.

\section{Conclusion}

The study equations in conservative form for simulating the dynamics of nonlinear waves on the liquid film flowing down a vertical plane has been performed. It has been found that in the computational domain ${\eta ' \in [ - 1,1]}$ extended along the transverse coordinate the equations (\ref{eq:equations}) with boundary conditions (\ref{eq:bound1})-(\ref{eq:bound2}) is invariant under parity transformations (\ref{eq:symmetry}). It was numerically proved that at moderate Reynolds numbers the steady-state travelling solutions of the problem (\ref{eq:equations})-(\ref{eq:bound2}) in the area, extended along the transverse coordinate, are characterized by the detected symmetry (\ref{eq:sym_bound1}). Relevance of this symmetry consideration for the numerical solution of the problem using spectral and pseudospectral methods has been demonstrated.

\bibliography{sources}

\end{document}